\documentclass[aps,prb,reprint, amsmath,amssymb,reprint,superscriptaddress]{revtex4-2}

\usepackage{mathrsfs,amsfonts} 
\usepackage{bbm}
\usepackage[colorlinks,citecolor=blue,linkcolor=red,urlcolor=blue]{hyperref}
\usepackage{color}
\usepackage{graphicx}
\usepackage{subfigure}
\usepackage{verbatim}
\usepackage{fourier}

\usepackage{graphicx}
\usepackage{dcolumn}
\usepackage{bm}

\begin{document}
\title{Casimir-Lifshitz force between graphene-based structures out of thermal equilibrium}

\author{Youssef Jeyar}
\email{youssef.jeyar@umontpellier.fr}
\affiliation{Laboratoire Charles Coulomb (L2C), UMR 5221 CNRS-University of Montpellier, F-34095 Montpellier, France}
\author{Kevin Austry}
\affiliation{Laboratoire Charles Coulomb (L2C), UMR 5221 CNRS-University of Montpellier, F-34095 Montpellier, France}
\author{Minggang Luo}
\affiliation{Laboratoire Charles Coulomb (L2C), UMR 5221 CNRS-University of Montpellier, F-34095 Montpellier, France}
\author{Brahim Guizal}
\affiliation{Laboratoire Charles Coulomb (L2C), UMR 5221 CNRS-University of Montpellier, F-34095 Montpellier, France}
\author{H. B. Chan}
\affiliation{Department of Physics, The Hong Kong University of Science and Technology, Clear Water Bay, Kowloon, Hong Kong, China}
\affiliation{William Mong Institute of
Nano Science and Technology, The Hong Kong University of Science and Technology, Clear Water Bay, Kowloon, Hong Kong, China}
\affiliation{Center for
Metamaterial Research, The Hong Kong University of Science and Technology, Clear Water Bay, Kowloon, Hong Kong, China}
\author{Mauro Antezza}
 \email{mauro.antezza@umontpellier.fr}
\affiliation{Laboratoire Charles Coulomb (L2C), UMR 5221 CNRS-University of Montpellier, F-34095 Montpellier, France}
\affiliation{Institut Universitaire de France, 1 rue Descartes, Paris Cedex 05, F-75231, France}

\date{\today}
\begin{abstract}
      We study the non equilibrium Casimir-Lifshitz force between graphene-based parallel structures held at different temperatures and in presence of an external thermal bath at a third temperature. The graphene conductivity, which is itself a function of temperature, as well as of chemical potential, allows us to tune in situ the Casimir-Lifshitz force. We explore different non equilibrium configurations while considering different values of the graphene chemical potential. Particularly interesting cases are investigated, where the force can change sign going from attractive to repulsive or where the force becomes non monotonic with respect to chemical potential variations, contrary to the behaviour under thermal equilibrium. 

\end{abstract}
\maketitle


\section{Introduction}

Van der Waals dispersion forces occur between polarisable objects and are a quantum, relativistic and macroscopic physical manifestation of both quantum and thermal fluctuations of the electromagnetic field. Generally, these forces increase rapidly as the separation between the objects decreases, becoming dominant over other forces at the micron scale and below. As a result, they could play an important role in nano- and micro-electromechanical systems. If the whole system is at thermal equilibrium, these forces are described by the Lifshitz, Dzyaloshinskii and Pitaevskii theory \cite{osti_4359646,Lifshitz_Pressure} which, in the middle of the 50's, considerably generalized the Casimir ideal-reflectors configuration \cite{Casimir48}. This allowed to  account for objects made by real materials and consider finite temperature, opening the possibility to study a large number of new experimental realizations. We hence name this general van der Waals interaction as Casimir-Lifshitz (CL) force. 
      Fifty years later, that theory has been further generalized to systems out of thermal equilibrium (OTE) where the different objects are held at different temperatures and are immersed in a thermal environmental bath held at a third temperature, with the full system being in a stationary OTE state. The new non-equilibrium theory was first developed for atom-surface configurations \cite{AntezzaPhD,AntezzaPRL2005, AntezzaJPA,PhysRevLett.97.223203}, showing rich and unexpected behaviors (change of sign of the force which can possibly become repulsive, different temperature and separation distance power law,...). Some of these predictions were verified in experiments with ultracold atoms trapped close to a hot substrate \cite{CornellExp,PhysRevA.70.053619}. Internal atomic non-equilibrium features have been also investigated \cite{PhysRevLett.100.253201,PhysRevA.79.032101}. This theory was subsequently generalized to macroscopic planar objects \cite{AntezzaPhD,AntezzaPRA2008}, and finally to (two or more) objects of arbitrary shape and dielectric function \cite{PhysRevA.80.042102,AntezzaPRA11,AntezzaEPL2011,AntezzaPRA2014}, opening new opportunities in the study of the CL interaction in the general and richer framework of non-equilibrium systems. A single closed-form expression based on scattering matrix theory has been derived \cite{AntezzaPRA11,AntezzaPRA2014} accounting for both CL force out of equilibrium and the concurrent radiative heat transfer \cite{PolderVanHove} occurring between the objects, hence unifying the momentum and energy transfer in those systems. Afterwards, expressions have been derived also using the Green's function formalism \cite{PhysRevLett.106.210404,PhysRevB.86.115423}.
      
\begin{figure}[ht!]
\centering
{\includegraphics[width=8.5cm]{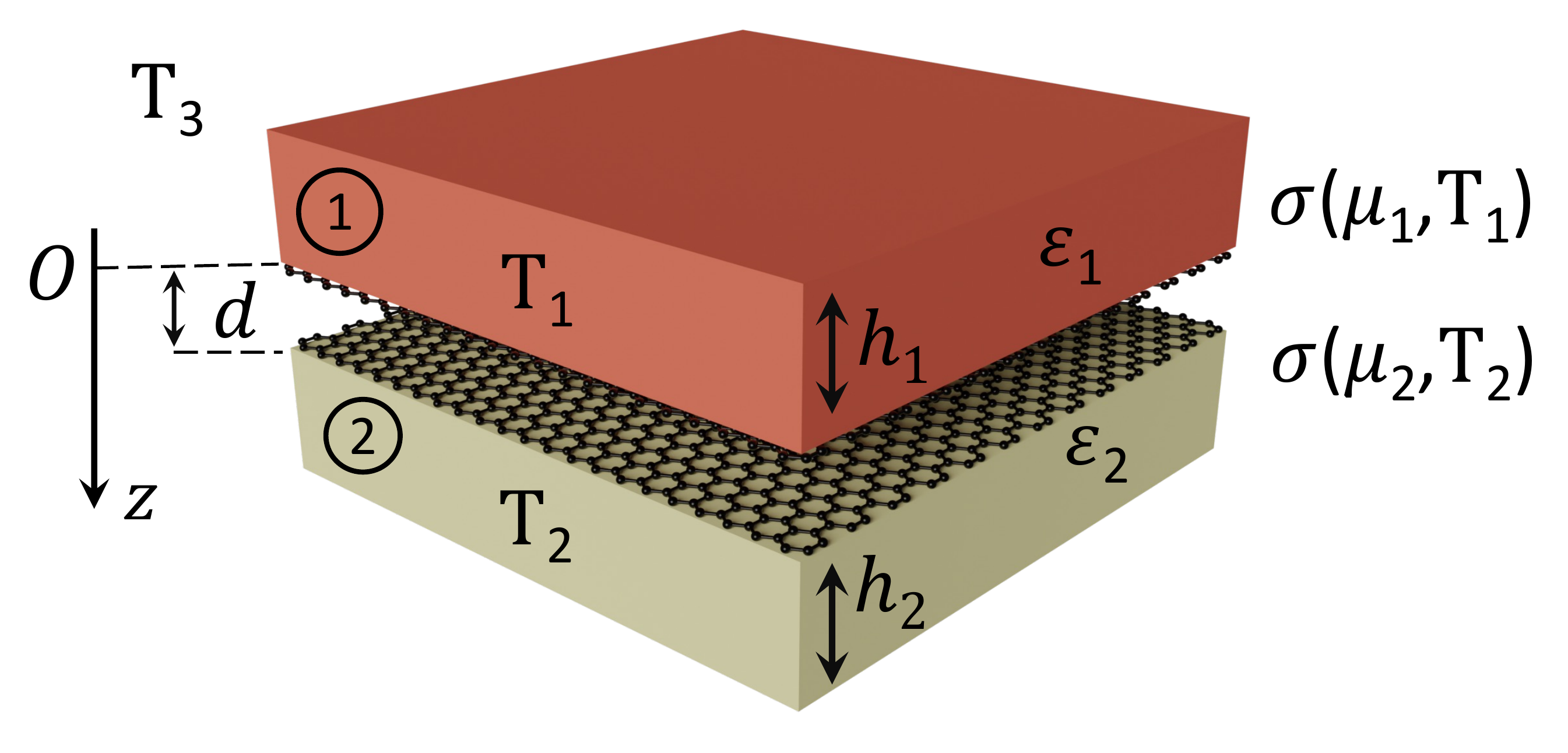}}
\caption{Sketch of the system under consideration, consisting of two parallel graphene-based structures $1$ and $2$ held at temperatures $T_1$ and $T_2$ respectively, in presence of an environmental bath at temperature $T_3$. The two graphene sheets have conductivity $\sigma_1(\omega,\mu_1,T_1)$ and $\sigma_2(\omega,\mu_2,T_2)$, depending on their chemical potential $\mu_i$ and temperature $T_i$.}
\label{Figure1}
\end{figure}
      
      With this general OTE theory at hand, the CL force OTE has been intensively explored for different geometric configurations (atom-surface \cite{AntezzaPRL2005,CornellExp,AntezzaPRA11,AntezzaPRA2014} parallel planes \cite{AntezzaPRA2008,AntezzaPRA11,AntezzaPRA2014}, diffraction gratings \cite{Noto}). In addition to geometric features, the CL force is also very sensitive to the material dielectric properties. It is hence interesting to study the effects that  emerging materials can bring to the features of CL force OTE. One of the most interesting materials today is graphene, due to its peculiar properties \cite{Svetovoy_2011,PhysRevB.87.205433}, like, for instance, the possibility to tune, in situ, its Fermi energy (or chemical potential $\mu$) and hence its electric conductivity  by a simple applied voltage, and the presence of thermally activated surface modes exalting the thermal CL force at very short separations \cite{PhysRevB.80.245424,PRL_Chahine,PRL_Mohideen}. By using the explicit atom-surface expression derived in \cite{AntezzaPRA11}, the atom-graphene force has been recently investigated \cite{MostepanenkoPRA2022} for $\mu=0$. In this paper we investigate the CL force OTE between two graphene-based parallel structures (see scheme on Fig.\ref{Figure1}) made by suspended graphene or SiO$_2$ slabs coated with graphene, in different thermal non-equilibrium configurations involving three different temperatures, and also for different values of chemical potential $\mu$.
      
In sec. \ref{Model} we describe the physical systems, the dielectric properties of the involved materials and the general model and formalism. In sec. \ref{Results} we present and analyse the numerical calculations of the force in different configurations.

\section{PHYSICAL SYSTEM and MODEL \label{Model}}
The physical system under consideration is depicted in Fig.\ref{Figure1}. It consists of two dielectric slabs of thicknesses $h_1$ and $h_2$ and relative dielectric permittivities $\varepsilon_1(\omega)$ and $\varepsilon_2(\omega)$ that are facing each other and separated with distance $d$. They are covered with graphene sheets having conductivities $\sigma_1(\omega,\mu_1,T_1)$ and $\sigma_2(\omega,\mu_2,T_2)$ respectively where $\mu_1$ (respectively $\mu_2$) is the chemical potential of graphene sheet 1 (respectively graphene sheet 2) and $T_1$ (respectively $T_2$) is the temperature of graphene covered slab 1 (respectively graphene covered slab 2). $T_3$ is the temperature of the environment. 
 {We emphasize that the environmental temperature is not the temperature of the electromagnetic field. Instead, it is just the temperature of the walls surrounding the two bodies.}

\subsection{Casimir-Lifshitz Pressure}

The Casimir-Lifshitz pressure (CLP) out of thermal equilibrium between two parallel homogeneous dielectric slabs of finite thicknesses, covered with graphene sheets,  separated by a distance $d$ and acting on body 1 can be expressed as  [\onlinecite{AntezzaPRA2008}] 
\begin{equation}
\begin{split}
P_{1z}(T_1,T_2,T_3) =&\dfrac{P_z^{(eq)}(T_1)+P_z^{(eq)}(T_2)}{2}+\Delta_{1z}(T_1,T_2,T_3)
\label{Pgeneral}
\end{split}
\end{equation}
where all terms in this expressions depend on the chemical potential $\mu_1$ and $\mu_2$ of the two graphene layers. Here $P_z^{(eq)}$ stands for the CLP at thermal equilibrium and is given by 
\begin{equation}
\begin{split}
P_z^{(eq)}(T) =&\dfrac{k_BT}{\pi}\sum_{n=0}^{\infty }\sideset{'}{}\int_{0}^{\infty}dQ Q q \sum_p \left[ \dfrac{e^{2qd}}{\tilde{\rho}_{1p}\tilde{\rho}_{2p}}-1\right]^{-1}
\label{Pequilibrium}
\end{split}
\end{equation}
where the prime on the sum means that the $n$ = 0 term is divided by 2, $q = \sqrt{(\xi_n/c)^2 + Q^2}$ and $\xi_n$=2$\pi n k_B T/\hbar$ are the Matsubara frequencies.  $\tilde{\rho}_{1p/2p}$ are the frequency-rotated reflection coefficients of bodies 1 and 2 respectively, $p$ standing for either $s$ (transverse electric TE) or $p$ (transverse magnetic TM) polarisation. 
The second term in the first equation corresponds to the nonequilibrium contribution, and is given by 
\begin{equation}
\begin{split}
\Delta_{1z}&(T_1,T_2,T_3)=\\
&A^{\textnormal{(ew)}}(T_1)-A^{\textrm{(ew)}}(T_2)+B_1^{\textrm{(pw)}}(T_1)-B_1^{\textrm{(pw)}}(T_2)\\
&+B_2^{\textnormal{(pw)}}(T_3)-B_2^{\textrm{(pw)}}(T_1)+B_3^{\textrm{(pw)}}(T_3)-B_3^{\textrm{(pw)}}(T_2)
\end{split}
\end{equation}

with
\begin{widetext}
\begin{eqnarray}
A^{\textnormal{(ew)}}(T)&=&\dfrac{\hbar}{2\pi^2}\sum_p \int_{0}^{\infty} d\omega \int_{\frac{\omega}{c}}^{\infty} dk\ k \ \textrm{Im} \ k_z n(\omega,T)\dfrac{ \textrm{Im(}\rho_{1p} \rho^*_{2p}\textrm{)}}{\vert D_p \vert^2} e^{-2d \textrm{Im} k_z},\notag \\
B_1^{\textnormal{(pw)}}(T)&=&-\dfrac{\hbar}{4\pi^2}\sum_p \int_{0}^{\infty} d\omega \int_{0}^{\frac{\omega}{c}} dk\ k   \ k_z n(\omega,T)\dfrac{ \vert\rho_{2p} \vert^2 - \vert \rho_{1p} \vert^2 + \vert \tau_{1p} \vert^2 \textrm{(}1-\vert\rho_{2p} \vert^2 \textrm{)}}{\vert D_p \vert^2} \notag ,\\
B_2^{\textnormal{(pw)}}(T)&=&-\dfrac{\hbar}{4\pi^2}\sum_p \int_{0}^{\infty} d\omega \int_{0}^{\frac{\omega}{c}} dk\ k   \ k_z n(\omega,T)\left[  \dfrac{ \vert\tau_{1p} \vert^2 [1+ \vert\rho_{2p} \vert^2 ( 1- \vert\tau_{1p} \vert^2)] }{\vert D_p \vert^2}- \vert \rho_{1p} \vert^2 -  \textrm{2Re}\left( \dfrac{\rho^*_{1p}\rho_{2p}\tau^2_{1p}}{D_p} e^{2ik_z(d+h_1)}\right)\right] \notag , \\
B_3^{\textnormal{(pw)}}(T)&=&-\dfrac{\hbar}{4\pi^2}\sum_p \int_{0}^{\infty} d\omega \int_{0}^{\frac{\omega}{c}} dk\ k   \ k_z n(\omega,T)\left [ \dfrac{ \ \vert \tau_{2p} \vert^2}{\vert D_p \vert^2} (1+ \vert\rho_{1p} \vert^2 -  \vert\tau_{1p} \vert^2 ) -1 \right],
\end{eqnarray}
\end{widetext}
where  
\begin{equation}
 D_{p} (\omega,\mu_1,T_1,\mu_2,T_2)=1-\rho_{1p}\rho_{2p}e^{2ik_zd} 
\end{equation}
and 
\begin{equation}
n(\omega,T) = \dfrac{1}{e^{\hbar\omega/k_BT}-1}.
\end{equation}
For TE polarisation, and for the configuration described above, the Fresnel reflection and transmission coefficients are given by \cite{PRL_Chahine,PhysRevB.95.245437},
\begin{equation}
\begin{split}
\rho_{j,s}(\omega,\mu_j,T_j) =&\dfrac{r_{a,j} +(1+r_{a,j} +r_{b,j})r_{c,j} e^{2ik_{z,j}h_j}}{1-r_{b,j}r_{c,j}e^{2ik_{z,j}h_j}}\\
\tau_{j,s}(\omega,\mu_j,T_j) =&\dfrac{t_{a,j} t_{c,j} e^{i(k_{z,j}-k_z)h_j}}{1-r_{b,j}r_{c,j}e^{2ik_{z,j}h_j}}
\end{split}
\end{equation}
where 
\begin{equation}
\begin{split}
r_{a,j}& = \dfrac{k_{z} -  \sigma_j\omega\mu_0-k_{z,j}  }{k_{z} +  \sigma_j\omega\mu_0+k_{z,j} }\\
r_{b,j}& = \dfrac{k_{z,j} -  \sigma_j\omega\mu_0-k_{z}  }{k_{z,j} +  \sigma_j\omega\mu_0+k_{z} }\\
r_{c,j}& = \dfrac{k_{z,j} -k_{z}  }{k_{z,j} +k_{z} }\\
\end{split}
\end{equation}
and
\begin{equation}
\begin{split}
t_{a,j}& = \dfrac{2k_z}{k_{z}+k_{z,j}+\mu_0\omega\sigma_j}\\ 
t_{c,j}& = \dfrac{2k_{z,j}}{k_z+k_{z,j}}\\ 
\end{split}
\end{equation}

Here, $j$=1,2 denotes slab 1 or slab 2,  $k_z= \sqrt{(\omega/c)^2 -Q^2}$ and $k_{z,j}= \sqrt{\varepsilon_j(\omega/c)^2 -Q^2}$ are the z-components of the wavevectors for vacuum and for body $j$ respectively. $\sigma_j$ is the graphene conductivity for each sheet and $\mu_0$ is the permeability of vacuum.\\

For TM polarisation, the Fresnel coefficients are given by \cite{PRL_Chahine,PhysRevB.95.245437}

\begin{equation}
\begin{split}
\rho_{j,p} =&\dfrac{r_{d,j} +(1-r_{d,j} -r_{e,j})r_{f,j} e^{2ik_{z,j}h_j}}{1-r_{e,j}r_{f,j}e^{2ik_{z,j}h_j}}\\
\tau_{j,p} =&\dfrac{t_{d,j} t_{f,j} e^{i(k_{z,j}-k_z)h_j}}{1-r_{e,j}r_{f,j}e^{2ik_{z,j}h_j}}
\end{split}
\end{equation}
%
%
where
\begin{equation}
\begin{split}
r_{d,j}& = \dfrac{\varepsilon_j k_{z} -  k_{z,j}  + \frac{\sigma_j}{\omega \varepsilon_0}k_{z} k_{z,j}}{\varepsilon_j k_{z} +  k_{z,j}  + \frac{\sigma_j}{\omega \varepsilon_0}k_{z} k_{z,j}}\\
r_{e,j}& = \dfrac{ k_{z,j}  -\varepsilon_j k_{z}  + \frac{\sigma_j}{\omega \varepsilon_0}k_{z} k_{z,j}}{  k_{z,j}  +\varepsilon_j k_{z} +  \frac{\sigma_j}{\omega \varepsilon_0}k_{z} k_{z,j}}\\
r_{f,j}& = \dfrac{ k_{z,j}  -\varepsilon_j k_{z}  }{  k_{z,j}  +\varepsilon_j k_{z} }\\
\end{split}
\end{equation}
 and
\begin{equation}
\begin{split}
t_{d,j}& = \dfrac{2k_z\varepsilon_j}{k_{z,j}+k_z\varepsilon_j+k_zk_{z,j}\frac{\sigma_j}{\omega\varepsilon_0}},\\ 
t_{f,j}& = \dfrac{2k_{z,j}}{k_z\varepsilon_j+k_{z,j}},\\ 
\end{split}
\end{equation}

with $\varepsilon_0$ being the vacuum permittivity.  \\

Finally, the reflection coefficients for imaginary frequencies can be obtained merely by setting $\omega=i\xi$, that is $\tilde{\rho}(\xi)=\rho(\omega=i\xi)$.

\subsection{Graphene conductivity and dielectric function}
The graphene conductivity $\sigma(\omega)$ for real frequencies can be written as a sum of an interband and an intraband contributions, i.e, $\sigma = \sigma_{\textnormal{inter}} +\sigma_{\textnormal{intra}}$ respectively given by [\onlinecite{Falkovsky_Graphene_1},\onlinecite{Falkovsky_Graphene_2},\onlinecite{Awan_Graphene_3}]:

 \begin{eqnarray} 
  \sigma_{\textnormal{intra}} (\omega) &=&  \dfrac{i8 \sigma_0  k_B T}{ \pi(\hbar\omega + i\hbar/\tau)} \ln\left[ 2 \cosh\left(\dfrac{\mu}{2k_BT}\right) \right], \;\;\; \\
 \sigma_{\textnormal{inter}} (\omega)&=&  \sigma_0\left[G \left( \dfrac{\hbar \omega}{2}\right) + i \dfrac{4\hbar \omega}{\pi}\int_0^{+\infty} \dfrac{G(x)-G(\frac{\hbar \omega}{2})}{(\hbar \omega)^2 - 4x^2} dx \right].\;\;\;\;\;
\end{eqnarray} 
\\
Here, $\sigma_0={e^2}/{4\hbar}$  {(expressed in Siemens)}, $e$  is the electron charge, $G(x) = \sinh (x/k_B T) / [\cosh (\mu/k_B T) + \cosh(x/k_B T)]$, T is the graphene sheet temperature, $\tau$ the relaxation time (we use $\tau = 10^{-13}s$) and $\mu$ the chemical potential.  {This model is accurate enough for the range of distances we study. In case of very short separations of very few nanometers more sophisticated models \cite{G1} can be used to obtain an even more precise estimation of the force.} \\
\begin{figure}[ht!]
\centering
{\includegraphics[width=9cm]{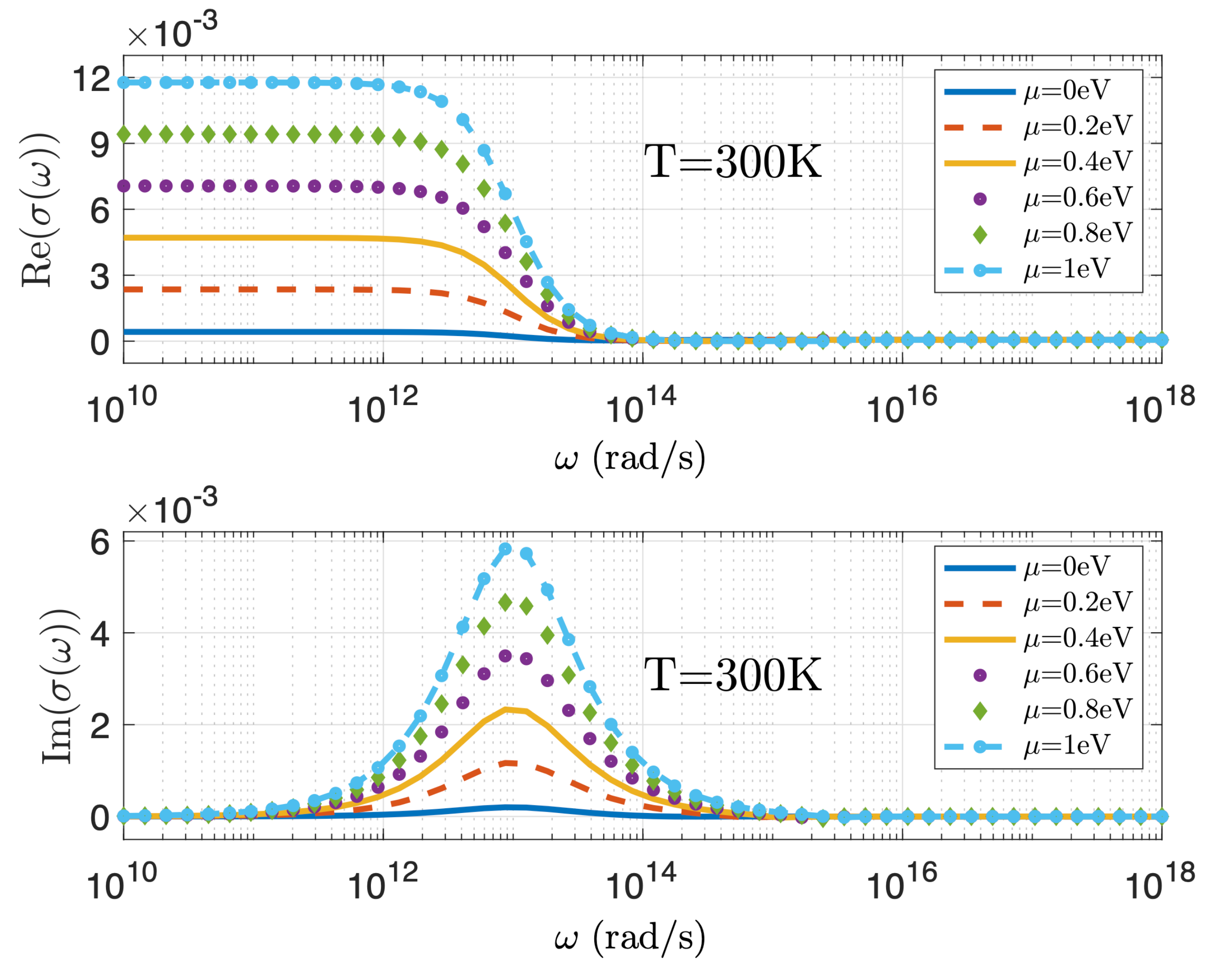}}
\caption{Graphene conductivity (real and imaginary part) at real frequencies for several values of $\mu$ and at T=300K.}
\label{Figure2}
\end{figure}
\begin{figure}[ht!]
\centering
{\includegraphics[width=9cm]{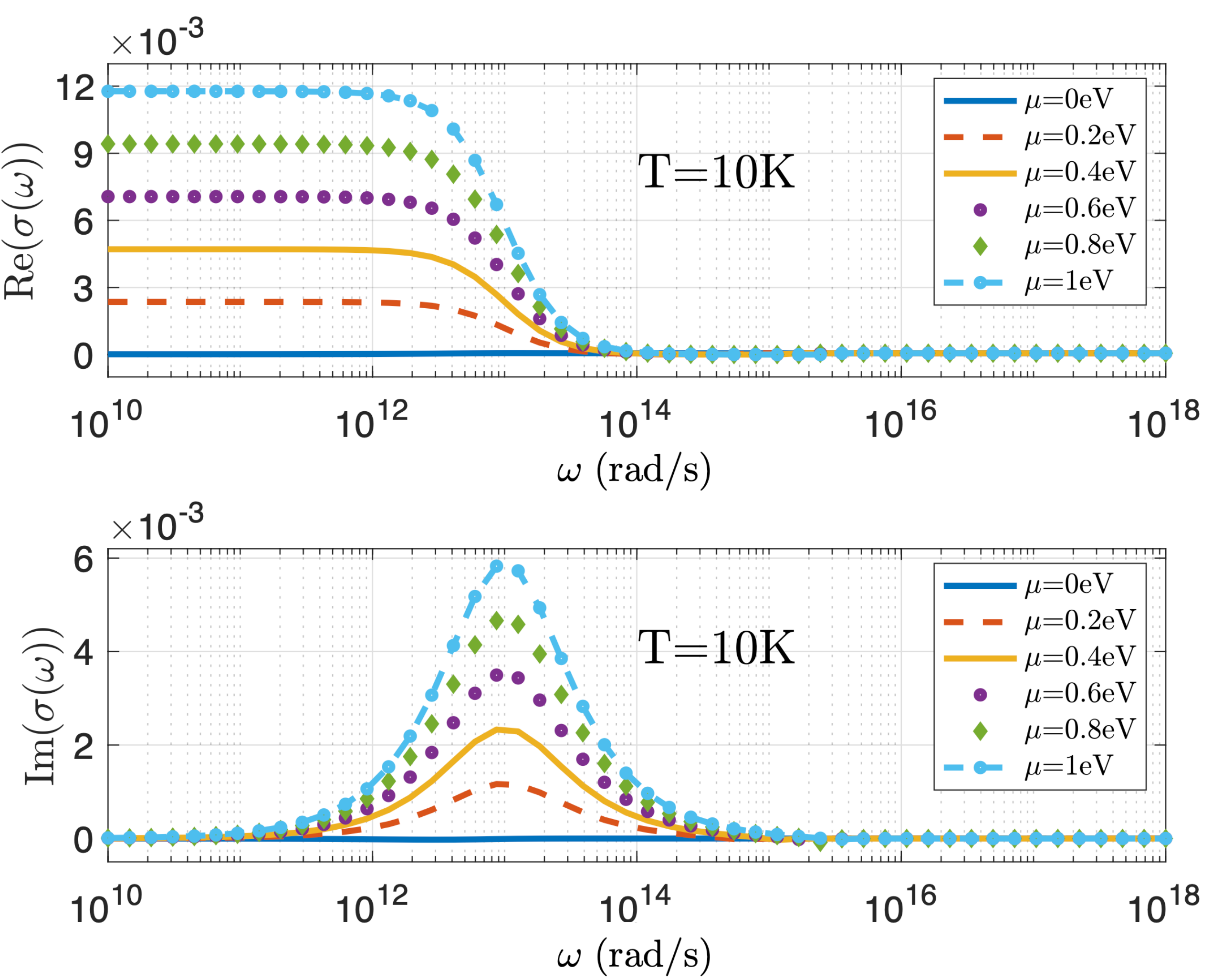}}
\caption{Graphene conductivity (real and imaginary part) at real frequencies for several values of $\mu$ and at T=10K.}
\label{Figure3}
\end{figure}
For Matsubara frequencies the graphene conductivity components can be expressed as [\onlinecite{PRL_Chahine}]  

\begin{eqnarray} 
\sigma_{\textnormal{intra}} (i\xi_n) &=& \dfrac{8\sigma_0 k_B T}{\pi(\hbar \xi_n+\hbar/\tau)}\ln\left[ 2 \cosh\left(\dfrac{\mu}{2k_BT}\right) \right],\\
\sigma_{\textnormal{inter}}  (i\xi_n)&=& \dfrac{\sigma_0 4\hbar \xi_n}{\pi}\int_0^{+\infty} \dfrac{G(x)}{(\hbar \xi_n)^2 + 4x^2} dx.
\end{eqnarray} 

%
\begin{figure}[ht!]
\centering
{\includegraphics[width=9cm]{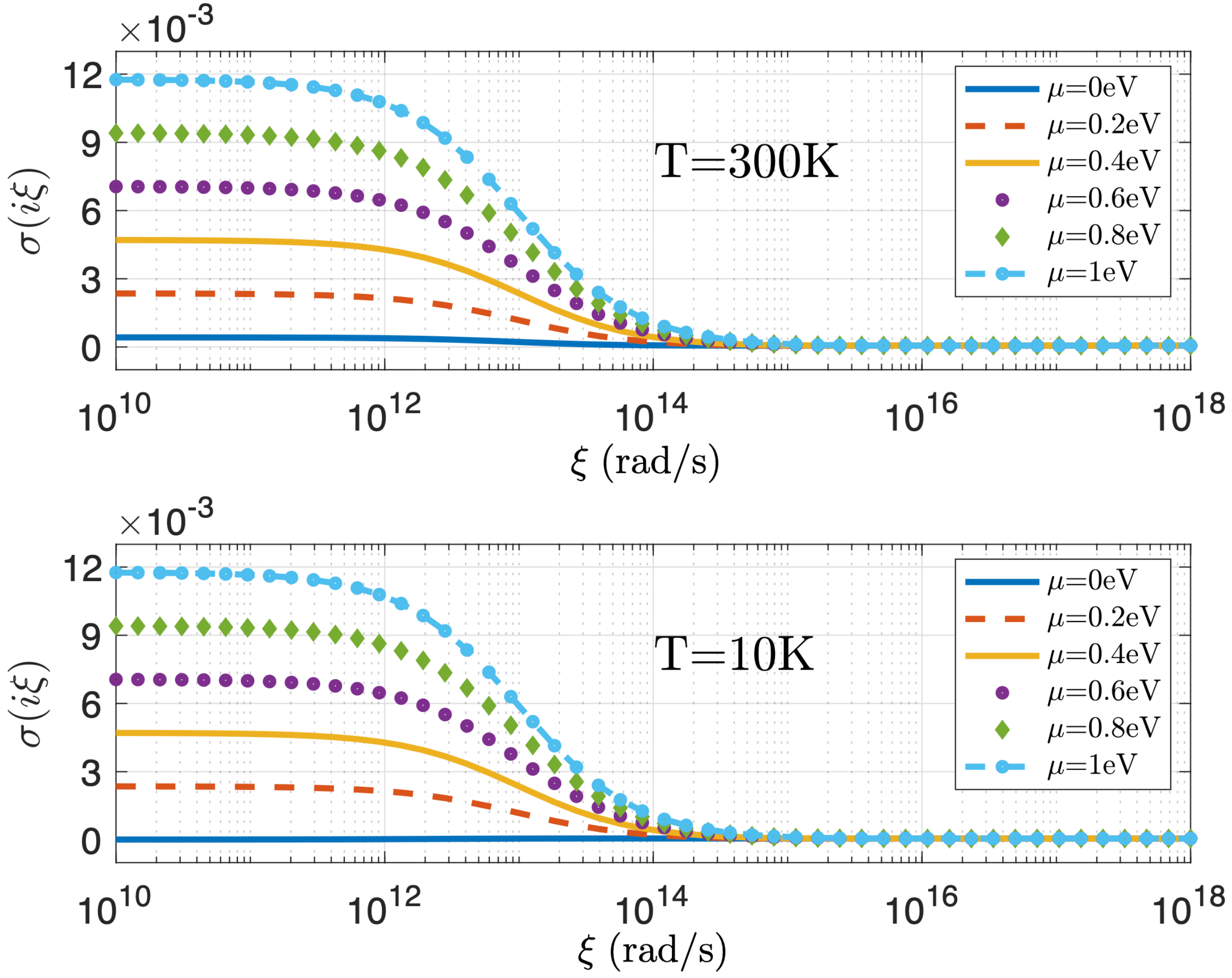}}
\caption{Graphene conductivity at imaginary frequencies for several values of $\mu$, at T=300K (upper panel) and T=10K (lower panel).}
\label{Figure4}
\end{figure}
\begin{figure}[ht!]
\centering
{\includegraphics[width=9cm]{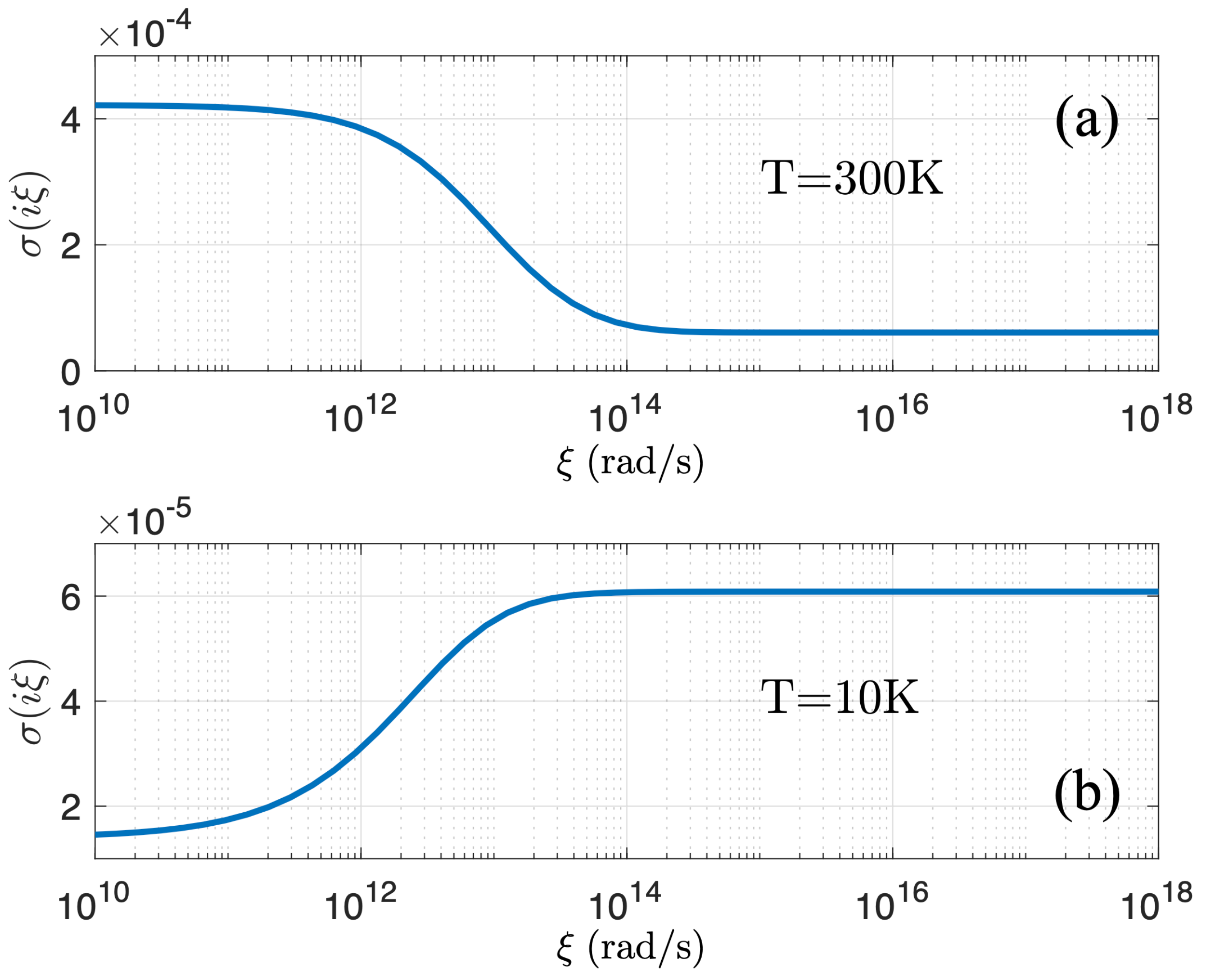}}
\caption{Graphene conductivity at imaginary frequencies for $\mu$=0 eV, at T=300K (upper panel) and T=10K (lower panel).}
\label{Figure5}
\end{figure}
\begin{figure}[ht!]
\centering
{\includegraphics[width=9cm]{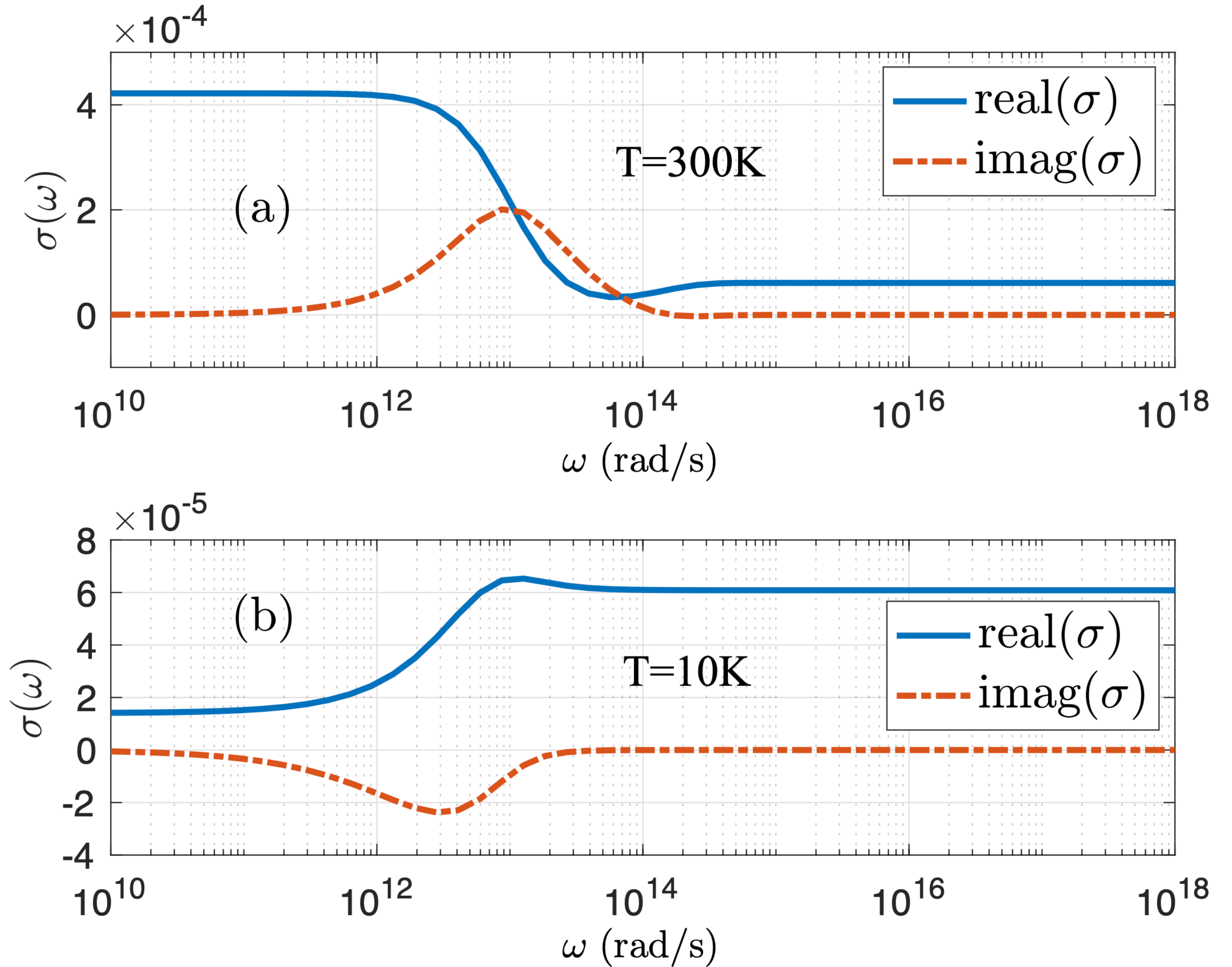}}
\caption{Graphene conductivity at real frequencies for $\mu$=0 eV, at T=300K (upper panel) and T=10K (lower panel).}
\label{Figure6}
\end{figure}

If we focus on the conductivity of graphene for T=10K and $\mu$=0 eV (as shown in Fig.\ref{Figure5}(b) for imaginary frequencies or Fig.\ref{Figure6}(b)  for real frequencies), we can clearly notice a different behaviour compared to T=300K and $\mu$=0 eV (Fig.\ref{Figure5}(a)) or T=10K and $\mu$ = 0.2 eV.
 \vspace*{1cm}

\begin{figure}[ht!]
\centering
{\includegraphics[width=9cm]{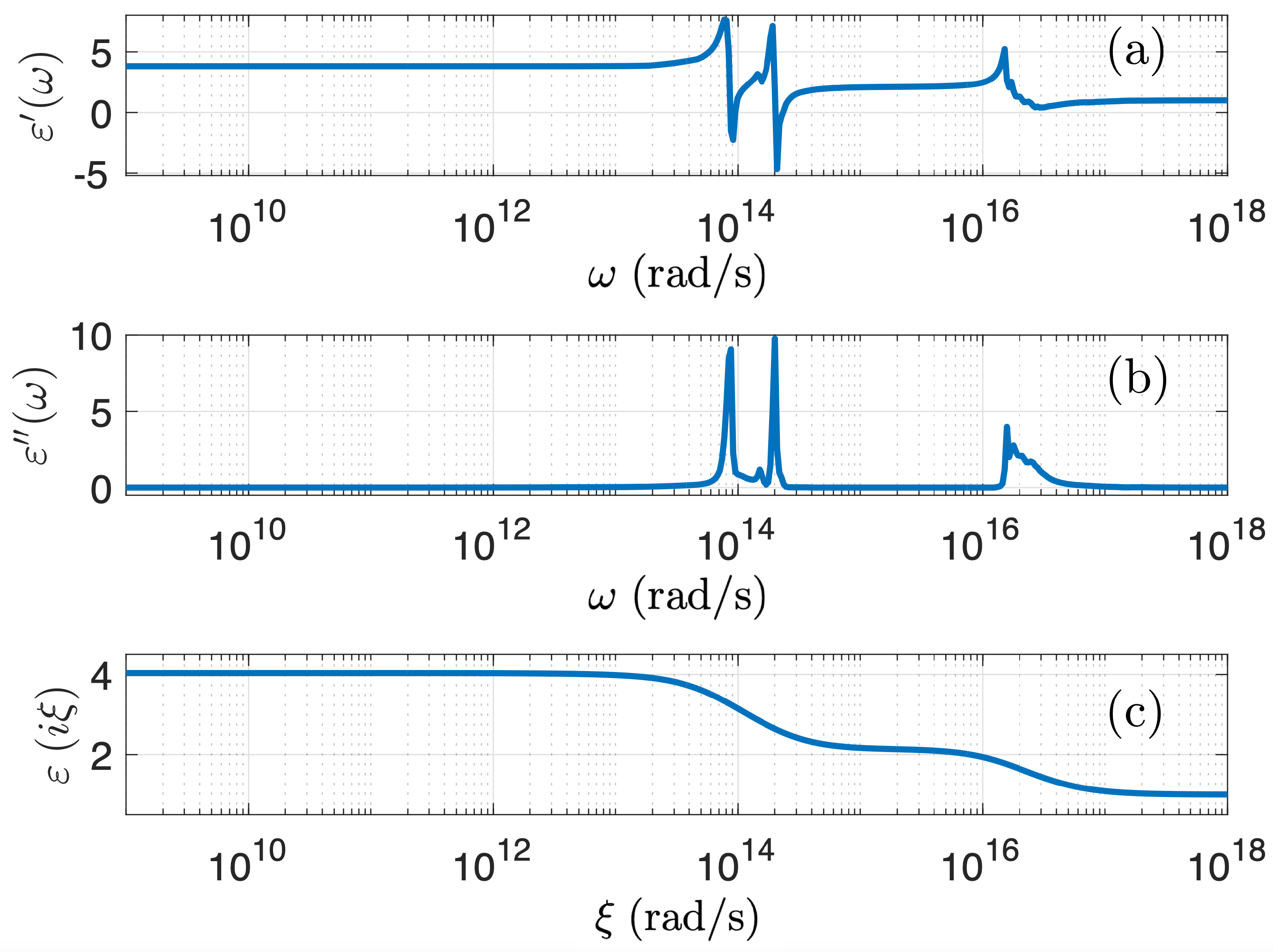}}
\caption{ {Relative} dielectric permittivity of fused silica (SiO$_2$). Real (a) and imaginary (b) parts of $\varepsilon(\omega)$ on real frequency axis, and (c) permittivity  $\varepsilon(i\xi)$ on the imaginary frequency axis.}
\label{Figure7}
\end{figure}

As already mentioned, we want to calculate the CLP out of thermal equilibrium between two parallel dielectric slabs covered with a graphene sheets. 
For our study, we choose fused silica (SiO$_2$) with relative dielectric permittivity  $\varepsilon(\omega)=\varepsilon'(\omega)+i\varepsilon''(\omega)$ taken from [\onlinecite{Book_SiO2}] .
In Fig.\ref{Figure7} we show the real part (a) and the imaginary part (b) of this dielectric function. For Matsubara frequencies, this function (which is necessarily real) can be deduced from the imaginary part $\varepsilon''(\omega)$ (for real frequencies) through the Kramers-Kronig relation  {given by the following equation}[\onlinecite{Kramers_Kronig_1},\onlinecite{Book_Lifshitz}]:

\begin{equation}
 \varepsilon(i\xi)=1+\dfrac{2}{\pi}\int_{0}^{\infty} \dfrac{\omega\varepsilon''(\omega)}{\omega^2+\xi^2}d\omega,
 \label{Kramers-Kronig}
 \end{equation}
 
  {represented in Fig.\ref{Figure7}(c).} 

\section{Results \label{Results}}

\subsection{Out of thermal equilibrium effect}

\begin{figure}[ht!]
\centering
{\includegraphics[width=9cm]{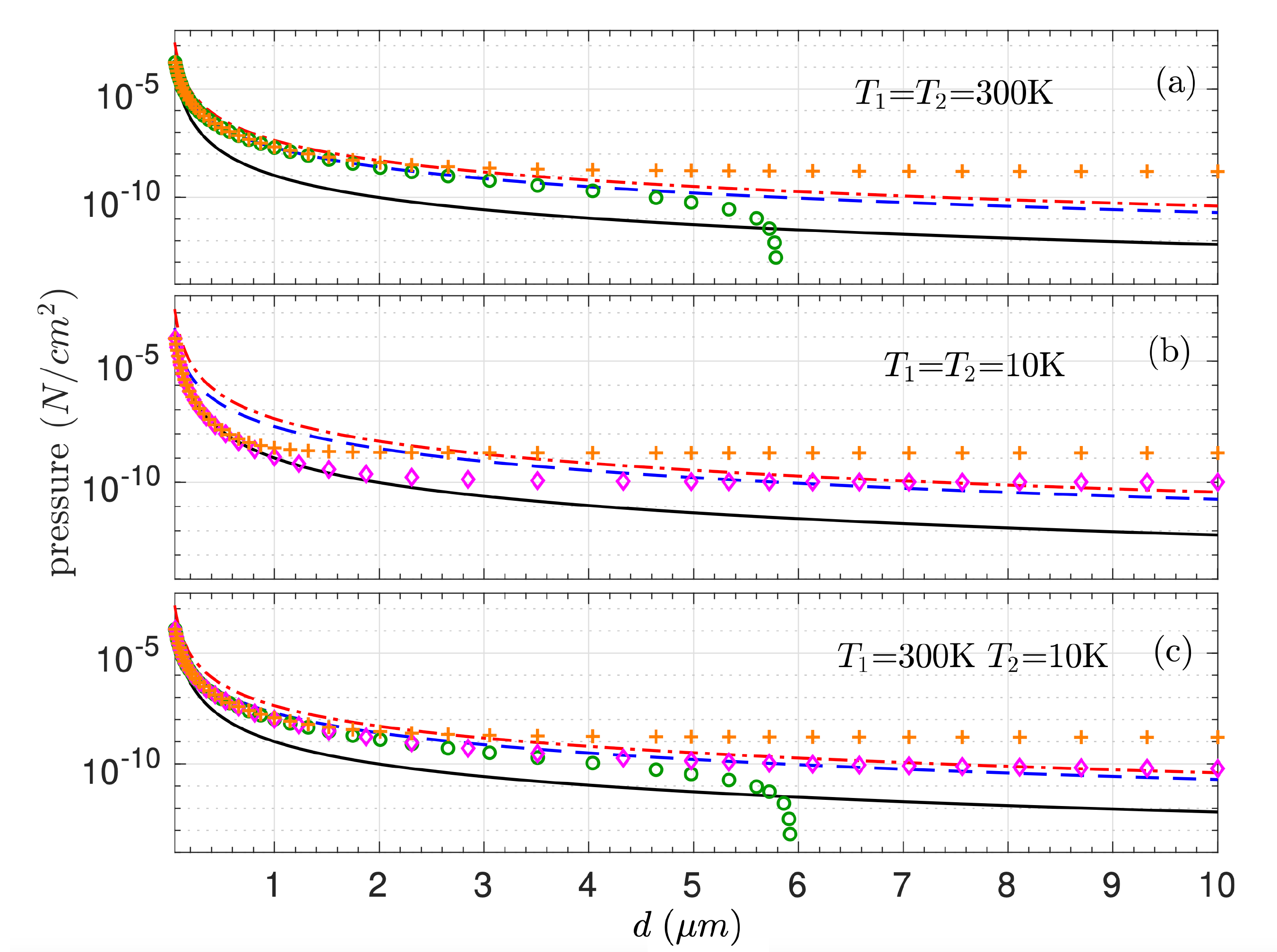}}
\caption{Comparison of the CLP out of thermal equilibrium [Eq. \eqref{Pgeneral}] and in thermal equilibrium [Eq. \eqref{Pequilibrium}] acting on body 1, between two graphene sheets with chemical potentials $\mu_1$=$\mu_2$=0 eV. Lines : equilibrium pressures at $T=10$ K (black solid), 300 K (blue dashed), and 600 K (red dash-dotted). Symbols: nonequilibrium pressures at T$_3$ =10 K (green circles), 300 K (magenta diamonds), and 600 K (orange plus). T$_1$=T$_2$ =300K in (a), T$_1$=T$_2$=10K in (b), and T$_1$=300K and T$_2$=10K in (c).}
\label{Figure8}
\end{figure}

In the previous section, we presented the theory behind the CLP out of thermal equilibrium and its dependence on temperature, the chemical potential of the graphene sheet and also the distance between objects. In this section, we will discuss the specific results obtained for the CLP between two graphene sheets ( $\mu_1$=$\mu_2$ = 0 eV) without a substrate.
The results are shown in  Fig.\ref{Figure8} , which displays the CLP as a function of distance for various configurations (see caption of Fig.\ref{Figure8} for details).\\

\begin{figure}[ht!]
\centering
{\includegraphics[width=9cm]{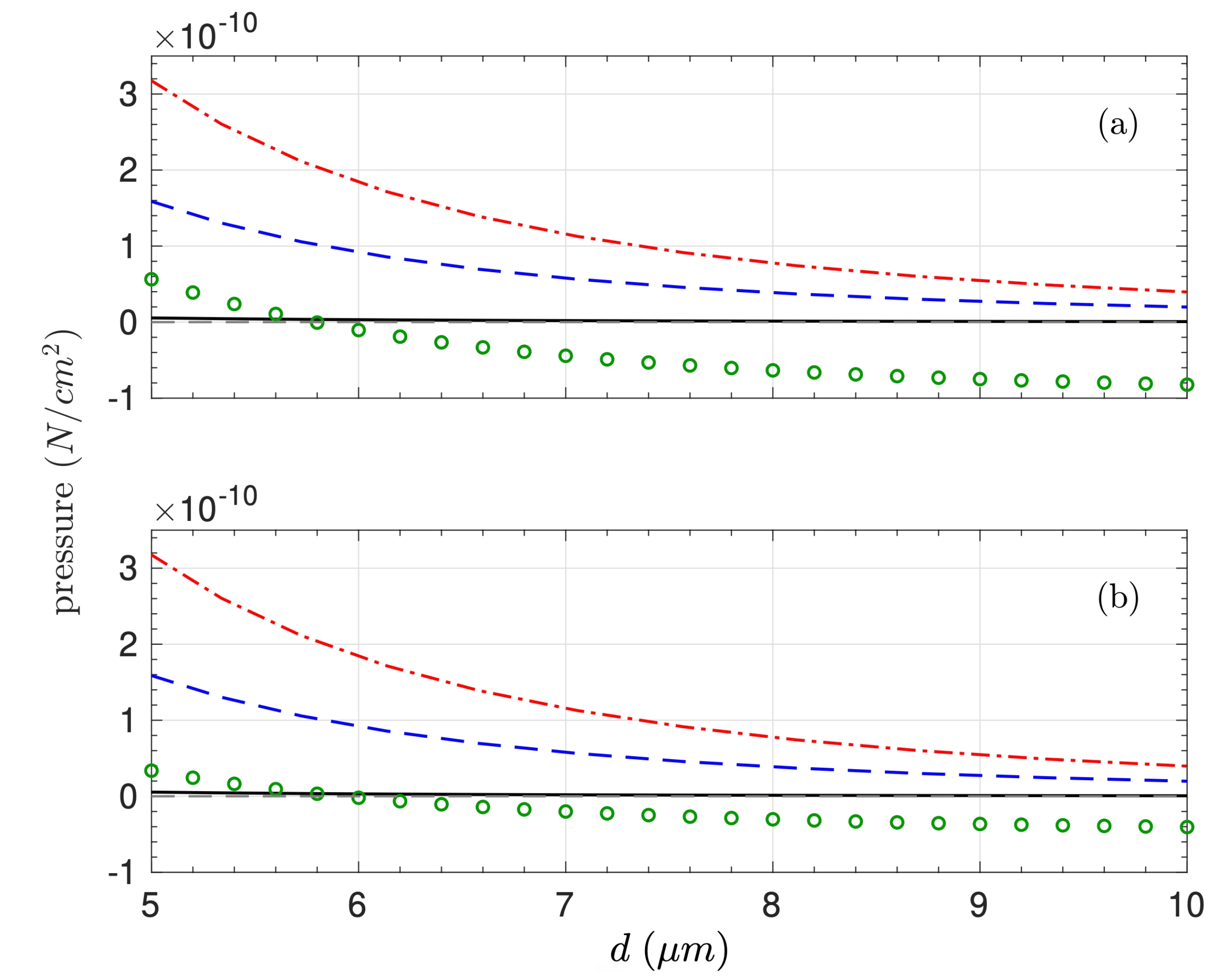}}
\caption{Zoom of Fig.\ref{Figure8} on linear scales, using the same conventions. The change in sign of the force is clearly visible}
\label{Figure9}
\end{figure}

The green circles in Fig.\ref{Figure8}(a)  show that when the two graphene sheets are at the same temperature $T_1$=$T_2$=300K, and the environment is at $T_3$=10K, the CLP is attractive at small $d$, causing them to be pushed towards each other. However, as the distance increases, the force decreases and eventually becomes repulsive at distances greater than 5.8$\mu m$. To better show this sign change, the pressure is displayed in linear scale (between 5 $\mu$m and 10 $\mu$m)  in Fig.\ref{Figure9}(a). \\

Considering a scenario where the first graphene sheet is at  $T_1$ = 300 K and the second at  $T_2$ = 10 K, the sign change in pressure occurs at a slightly larger separation distance of around 6 $\mu m$ (see Fig.\ref{Figure9}(b) with linear scale) compared to the previous case where  $T_1$=$T_2=300$K. Figure \ref{Figure8}(a) plots the CLP for the case when the environment is  at $T_3=600$K and the sheets are at the same temperature $T_1 = T_2 =300$K. As the CLP decreases with the separation distance, it becomes greater than the pressure at any thermal equilibrium (10K, 300K and 600K) for large separation distances (from around 3 $\mu$m). This behavior is also observed if we cool down one of the sheets ($T_1=300$K and $T_2=10$K, see Fig.\ref{Figure8}(c)) or both of them ($T_1 = T_2 =10$K, see Fig.\ref{Figure8}(b)). Such behavior is in contrast to the equilibrium case where the CLP decreases when the system is cooled down. It is worth noticing that the distance at which OTE-CLP becomes predominant is shifted by an amount that depends on the temperatures of the graphene sheets.        
For the system parameters we considered, this behavior appears to be a general trend happening whenever departing from an equilibrium situation and cooling down the sheets (while preserving the temperature of the environment).

\begin{figure}[ht!]
\centering
{\includegraphics[width=9cm]{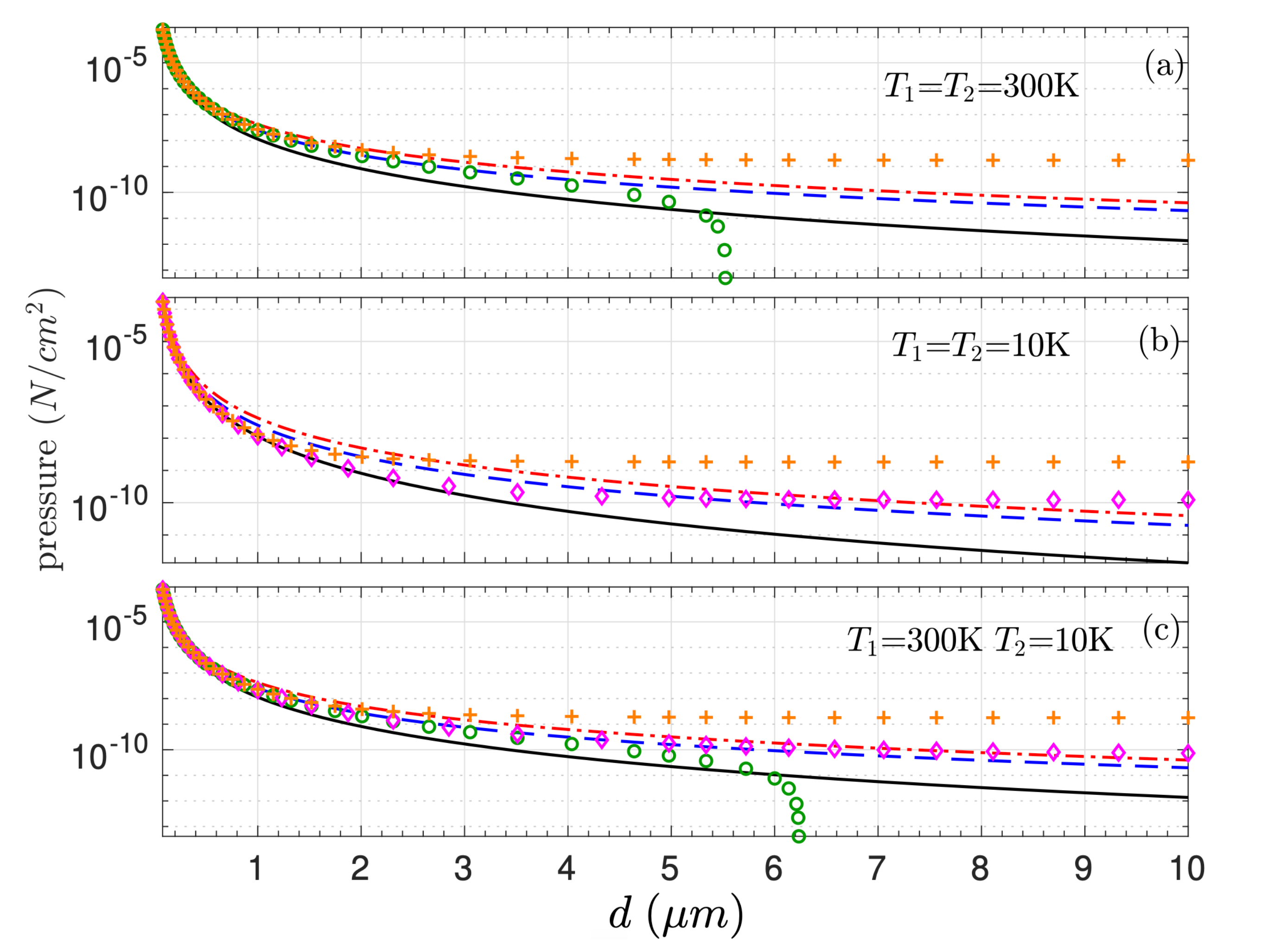}}
\caption{Comparison of the CLP out of thermal equilibrium [Eq. \eqref{Pgeneral}] and in thermal equilibrium [Eq. \eqref{Pequilibrium}] acting on body 1, between two fused silica slabs ($h_1$=$h_2$=2$\mu m$) covered with graphene sheets (chemical potential $\mu_1$=$\mu_2$=0eV). Lines : equilibrium pressures at $T=10$ K (black solid), 300 K (blue dashed), and 600 K (red dash-dotted). Symbols: non-equilibrium pressures at T$_3$ =10 K (green circles), 300 K (magenta diamonds), and 600 K (orange plus). T$_1$=T$_2$ =300K in (a), T$_1$=T$_2$=10K in (b), and T$_1$=300K and T$_2$=10K in (c).}
\label{Figure10}
\end{figure}

\begin{figure}[ht!]
\centering
{\includegraphics[width=8.5cm]{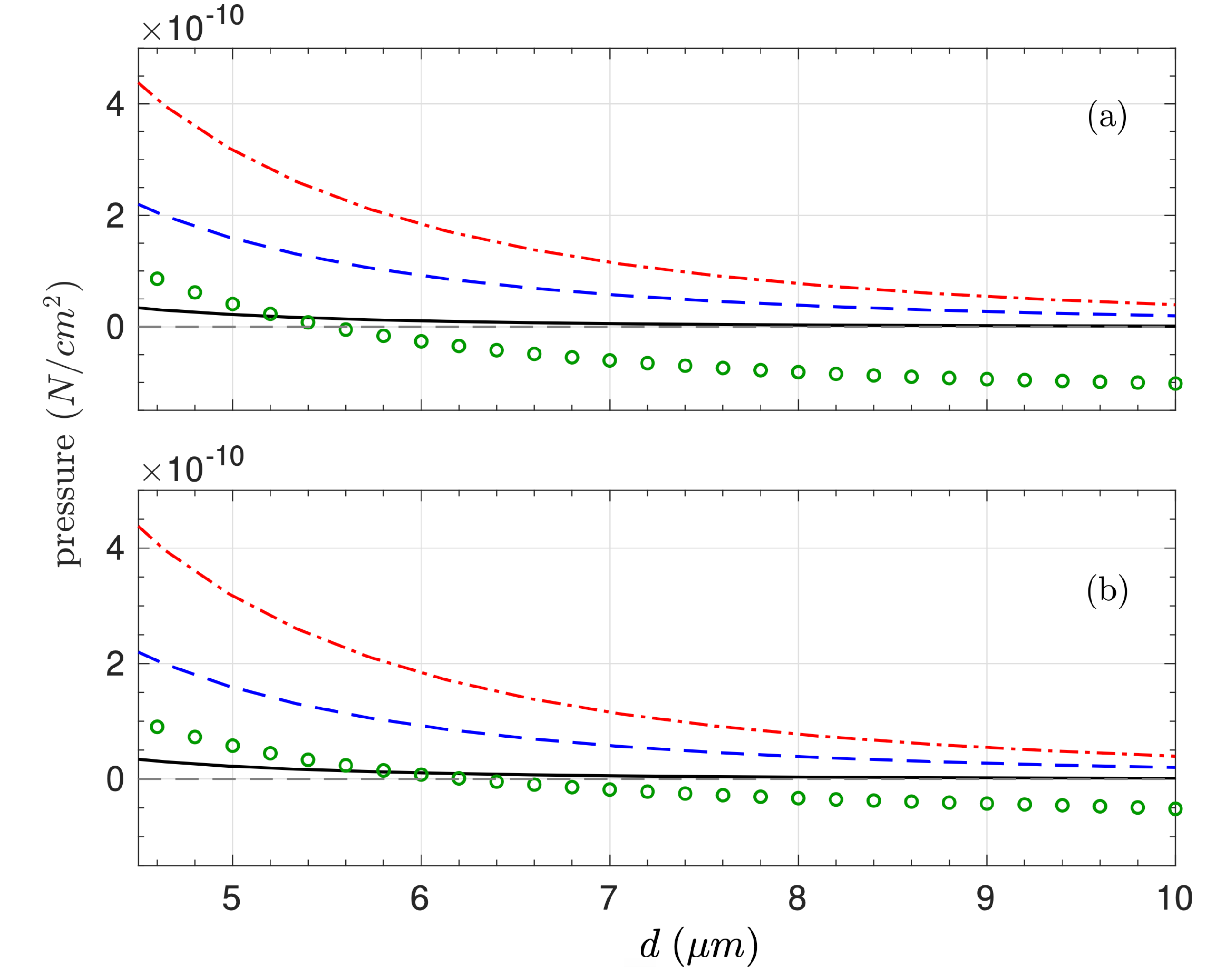}}
\caption{Zoom of Fig.\ref{Figure9} on linear scales, using the same conventions. The change in sign of the force is clearly visible}
\label{Figure11}
\end{figure}

Next, we extend our study to the case where two graphene sheets are placed on fused silica slabs with finite thicknesses  $h_1$ = $h_2$ = 2$\mu$m. The CLP is calculated and compared to the previous case of two graphene sheets without substrates.
Our results, shown in Fig.\ref{Figure10}, indicate that the behavior of the CLP shows similar features compared to the previous case. The attractive force is dominant at short distances and eventually becomes repulsive at larger distances when the environment is at $T_3$=10 K. However, we observe that the presence of the substrate slightly modifies the distance where the sign changes (As illustrated in Fig.\ref{Figure11}, which is plotted on a linear scale).  
Overall, the results suggest that the presence of the fused silica substrate does not lead to qualitatively new features for on the CLP between the graphene sheets.

\subsection{Chemical potential effect}

\subsubsection{Thermal equilibrium}
To begin, let us examine the chemical potential effect on the CLP at thermal equilibrium. To study the relative variation of the pressure as function of the chemical potential  (assuming that $\mu_1=\mu_2$) we define the function
\begin{equation}
\Delta P_\mu =\frac{P_{1z}(\mu)-P_{1z}(\mu=0)}{P_{1z}(\mu=0)},
\label{delta_P}
\end{equation}
where $P_{1z}$ is given by Eq. \eqref{Pequilibrium} or Eq. \eqref{Pgeneral}, for configurations at or out of thermal equilibrium, respectively.

\begin{figure}[h]
\centering
{\includegraphics[width=9cm]{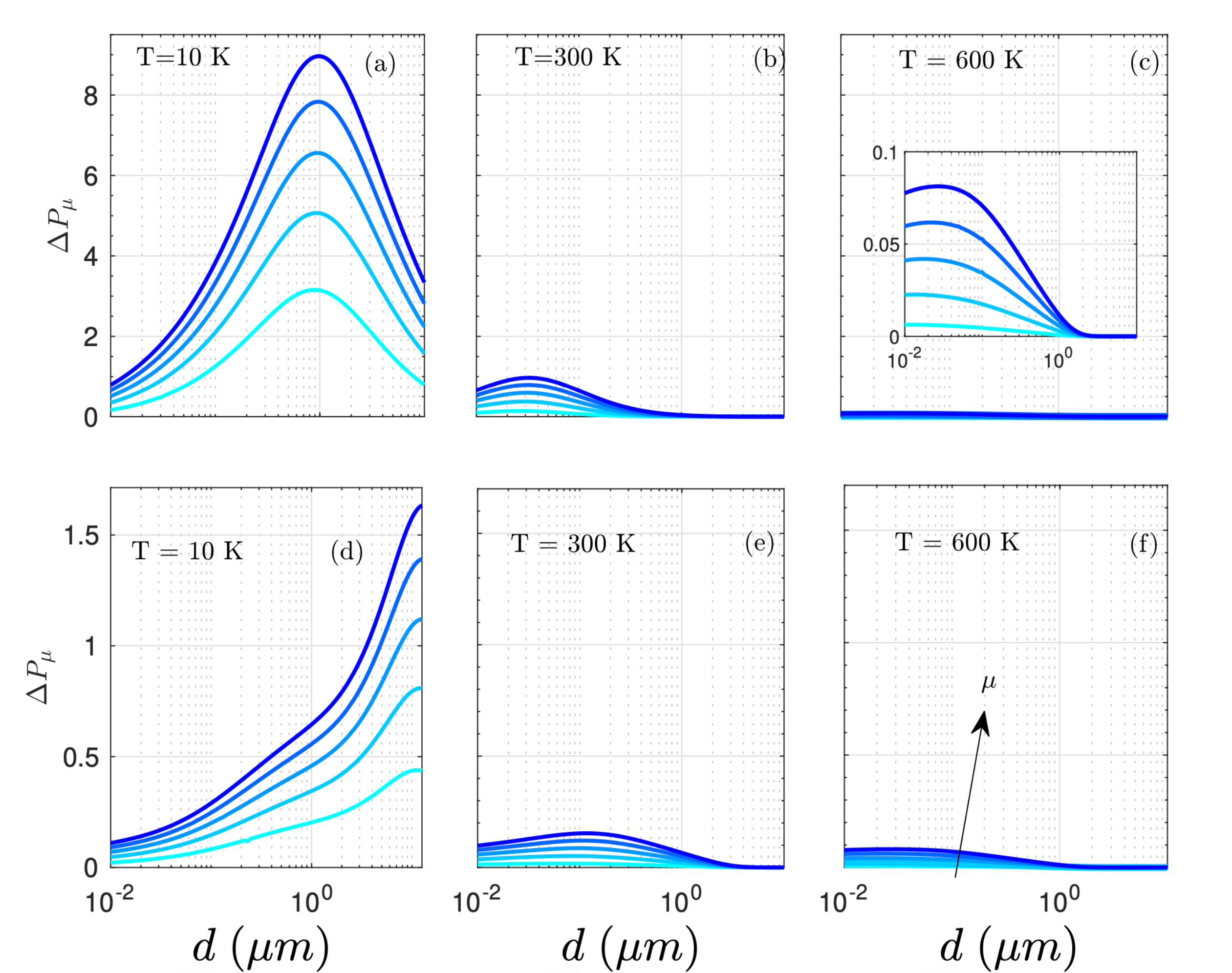}}
\caption{Relative variation $\Delta P_\mu$ of the CLP at thermal equilibrium as a function of the distance between the two suspended graphene(a,b,c) and covered slabs of $SiO_2$ (d,e,f). We plot $\Delta P_\mu $ for $\mu$=0.2, 0.4, 0.6, 0.8, 1 eV}
\label{Figure12_bis}
\end{figure}

Figure \ref{Figure12_bis} displays the relative variation $\Delta P_\mu$ for different values of the chemical potential with respect to the situation where $\mu$ = 0 eV.  This is done for two different configurations:  suspended graphene sheets (Fig.\ref{Figure12_bis}(a)(b)(c)) and slabs of $SiO_2$ covered with graphene sheets ((Fig.\ref{Figure12_bis}(d)(e)(f))).

Focusing on Fig.\ref{Figure12_bis}(a), which represents the T=10K case, it is evident that as the chemical potential increases, there is a simultaneous increase in the pressure. The graph exhibits a bell-shaped curve, with the relative variation being minimal at both small and large distances (universal graphene graphene interaction and the thermal limits  \cite{rodriguez2017casimir,woods2016materials}), and reaching its maximum at 1$\mu m$. Furthermore, when the temperature increases, the effect of the chemical potential diminishes significantly and the maximum of $\Delta P_\mu $ shifts towards smaller distances. For instance, at T=300K (Fig.\ref{Figure12_bis}(b)), the maximum variation occurs around 30 $n$m. This effect is due to the change in the thermal wavelength $\lambda_T$ which depends on the temperature. For the same reason, when the temperature is increased to T=600K (Fig.\ref{Figure12_bis}(c)),  $\Delta P_\mu $ becomes even smaller, and its maximum shifts to around 10 $n$m.

We then proceed to the second case involving two slabs of Si$O_2$ covered with graphene sheets. When T=10K (Fig.\ref{Figure12_bis}(d)), it can be observed that, compared to the case of two graphene sheets without a substrate, the maximum shifts to the right, because the CLP is dominated in this case by the $\mu$-independent substrate (for distances smaller than 8 $\mu$m), and the variation decreases. Furthermore, as the temperature increases, such as at T=300K (Fig.\ref{Figure12_bis}(e)) and T=600K (Fig.\ref{Figure12_bis}(f)), the maximum shifts towards the left, and the chemical potential effect  decreases.

\subsubsection{Out of thermal equilibrium : Suspended graphene sheets}
\begin{figure}[ht!]
\centering
{\includegraphics[width=9cm]{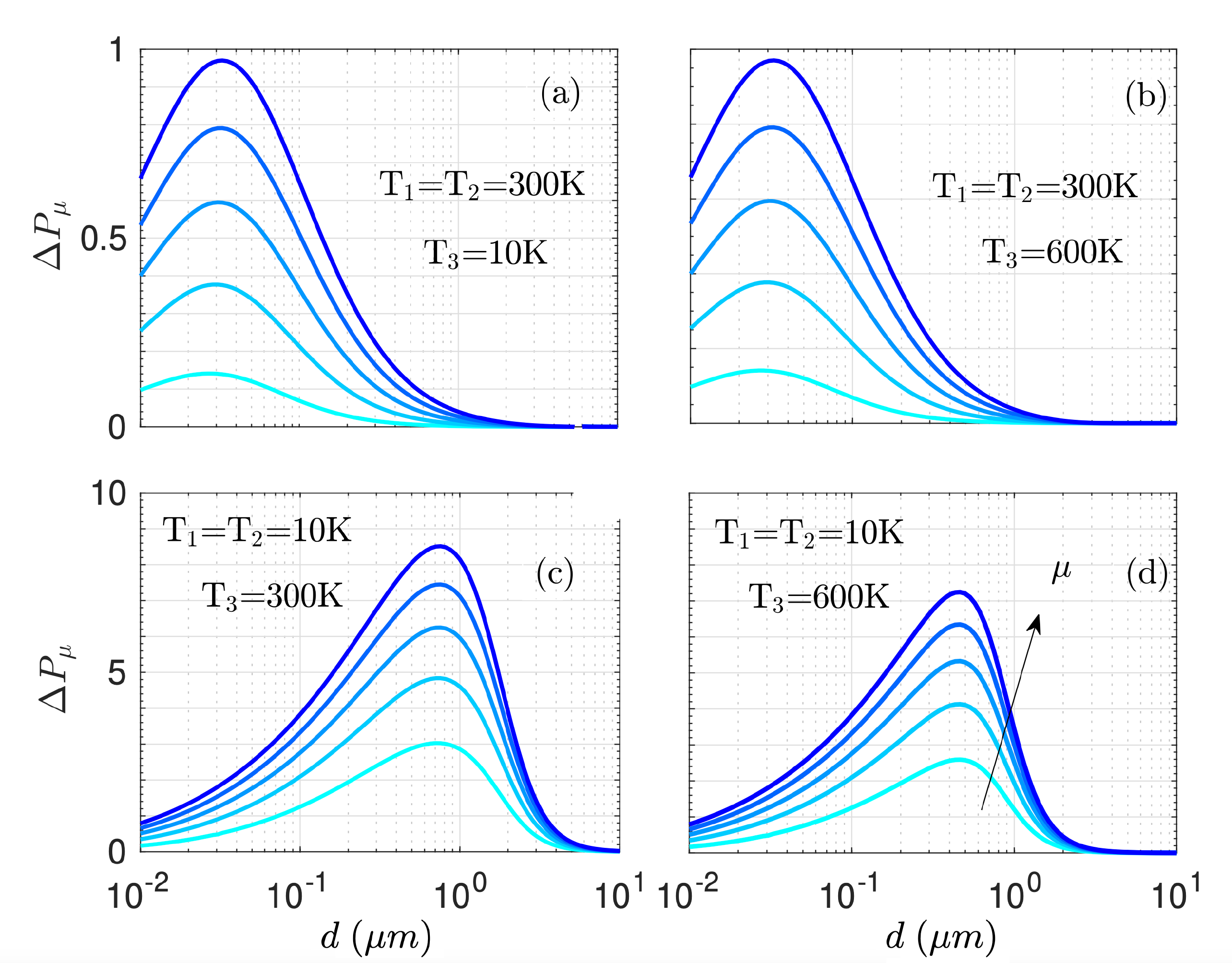}}
\caption{Relative variation $\Delta P_\mu$ of the CLP as a function of the distance between the two suspended graphene $T_1$=$T_2$.}
\label{Figure12}
\end{figure}

Besides examining the influence of non-equilibrium conditions on the CLP, we also explored how the chemical potential affects this pressure. Let us start considering the two graphene sheets being at the same chemical potential $\mu = \mu_1$=$\mu_2$. Fig.\ref{Figure12} shows $\Delta P_\mu $ between two graphene sheets for different values of the chemical potential (0.2, 0.4, 0.6, 0.8, 1 eV).
Our results reveal that when $T_1$=$T_2$=300K, varying the environment temperature $T_3$ does not lead to significant changes of the CLP for all $\mu $'s, as shown in Fig.\ref{Figure12}(a) and (b). This situation is comparable to the thermal equilibrium case as shown in Fig. \ref{Figure12_bis}(a).  In both \ref{Figure12}(a) and (b), as the chemical potential increases, the relative variation of the non-equilibrium pressure also increases, reaching approximately a value of 1 when $\mu=$1eV and the separation distance is near 30 $nm$. 
On the other hand, when  $T_1$=$T_2$=10K and $T_3$=300 K, as shown in Fig.\ref{Figure12}(c), the impact of the chemical potential becomes much stronger.  Specifically, when the separation distance is around 500 nm, the CLP with $\mu$= 1eV becomes about 10 times greater than when $\mu$= 0 eV and $d$ is near 500$n$m. 
The relative variation is further demonstrated in Fig.\ref{Figure12}(d) when the environment temperature is increased to 600K, where the effect is slightly lower than in the previous case. Our findings suggest that the chemical potential has a substantial impact only when the graphene sheets are at low temperatures.

\begin{figure}[ht!]
\centering
{\includegraphics[width=9cm]{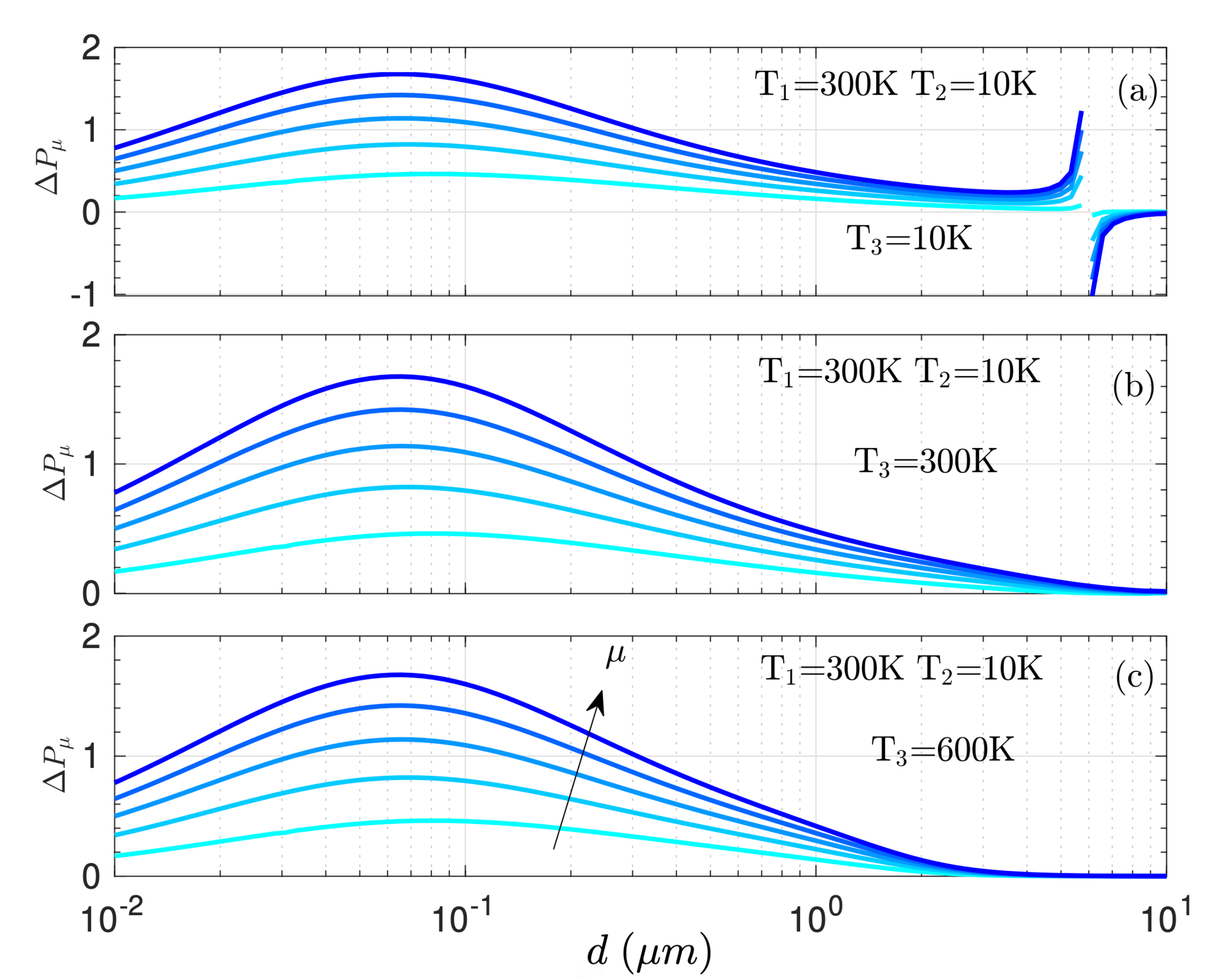}}
\caption{Relative variation $\Delta P_\mu$ of the CLP as a function of the distance between the two suspended graphene $T_1 \neq T_2$.}
\label{Figure13}
\end{figure}

In order to further understand the impact of the chemical potential on the CLP between two graphene sheets, we extend our investigation to the case where the two sheets are at different temperatures. As shown in Fig.\ref{Figure13}(a), where $T_1$ = 300K,  $T_2$ = 10K and $T_3$ = 10K, the relative variation increases with chemical potential, and the pressure for $\mu$=1eV is larger than that for $\mu$=0eV before the sign change. However, beyond a separation distance of 6$ \mu$m (where the pressure becomes repulsive) the CLP becomes smaller  when $\mu$=1 eV than when  $\mu$=0 eV.
Furthermore, it is worth noticing that the singularities in Fig.\ref{Figure13}(a) are due to the fact that we are dividing by the pressure at $\mu=0eV$, which is very low. We notice also that the relative variation of the pressure increases with the chemical potential, and is the most important when $\mu =$ 1 eV and $d = $ 60 nm,  except for the case where $T_3=$10K and $d > $ 6 $\mu$ m.    
\subsubsection{Out of thermal equilibrium : fused silica slabs covered with graphene sheets}

In the previous section, we examined the impact of the chemical potential on the (CLP) between two graphene sheets under non-equilibrium conditions.We now turn our attention to a system composed of two graphene sheets covering two fused silica substrates with identical thicknesses of $h_1 = h_2 = 2 \mu m$.

\begin{figure}[ht!]
\centering
{\includegraphics[width=9cm]{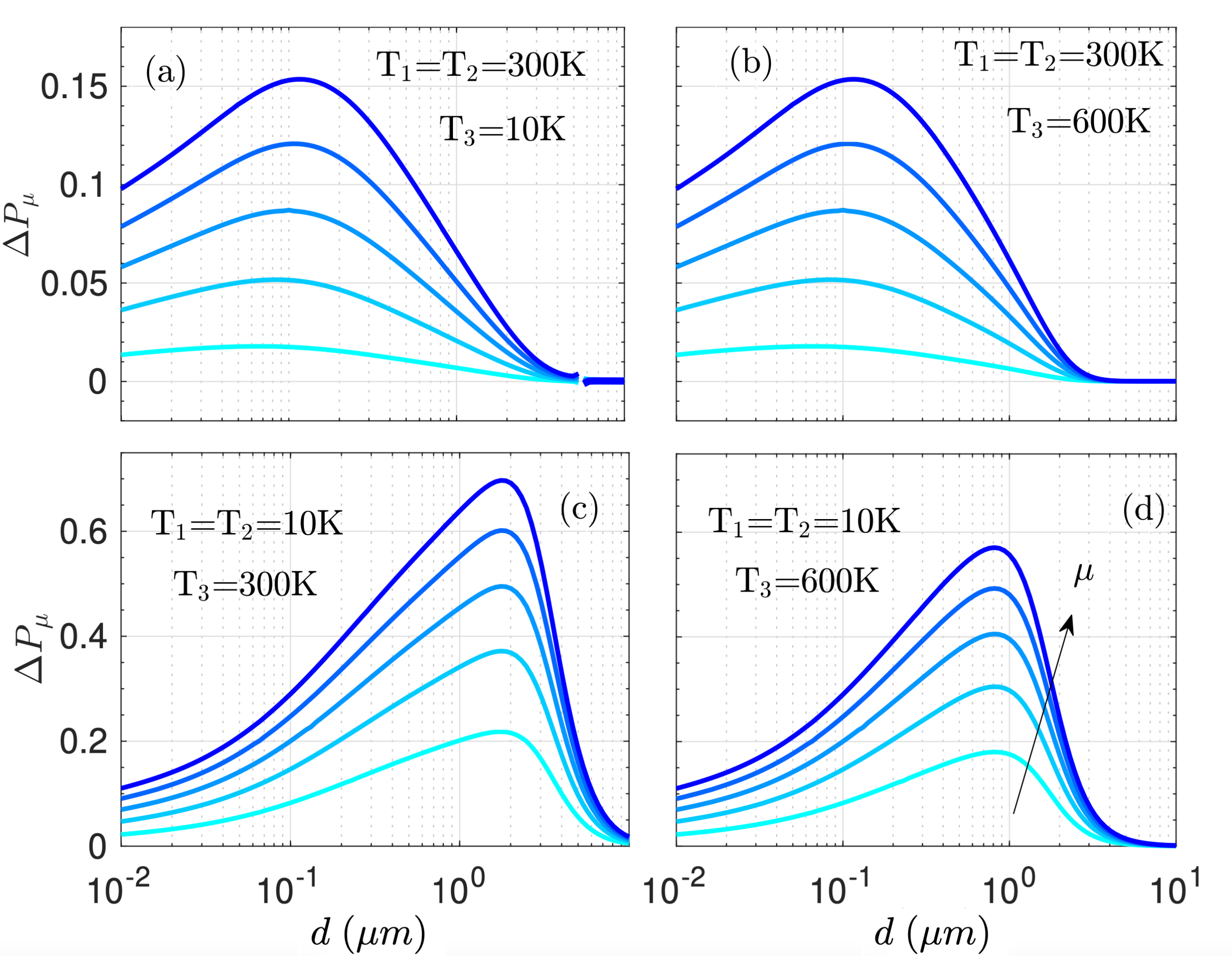}}
\caption{Relative variation $\Delta P_\mu$ of the CLP as a function of the distance between two slabs of $SiO_2$ covered with graphene sheets $T_1 = T_2$}
\label{Figure14}
\end{figure}

Our results, shown in Fig.\ref{Figure14}(a) and (b), reveal that the relative variation of the pressure in this system is significantly smaller than in the previous case, reaching a maximum value of only 0.15 when both objects are at 300K regardless of $T_3$. Furthermore, this maximum is shifted slightly to the right, towards a separation distance of 100$nm$ (30 $nm$ in the case of suspended graphene). 

It should be noted that when the two graphene sheets are at 300 K and the environment temperature is 10 K, the effect of the chemical potential after the change in sign, i.e., after 6$\mu$m, is very small.

On the other hand, as the temperature of the graphene sheets decreases to $T_1=T_2=10$ K, the relative variation increases, with a maximum value of 0.75 for $T_3=300$ K, as shown in Fig.\ref{Figure14}(c). In this case, we observe that the maximum occurs at a larger separation distance, between 1 $\mu$m and 2 $\mu$m, and shifts towards greater distances. 

Notably, for $T_3=600$ K, the maximum relative variation occurs between 900 $n$m and 1 $\mu$m, as shown in Fig.\ref{Figure14}(d), which is smaller than the previous case. Nevertheless, the maximum value is still significant, and the overall trend remains similar to the case of $T_3=300$ K.

\begin{figure}[ht!]
\centering
{\includegraphics[width=9cm]{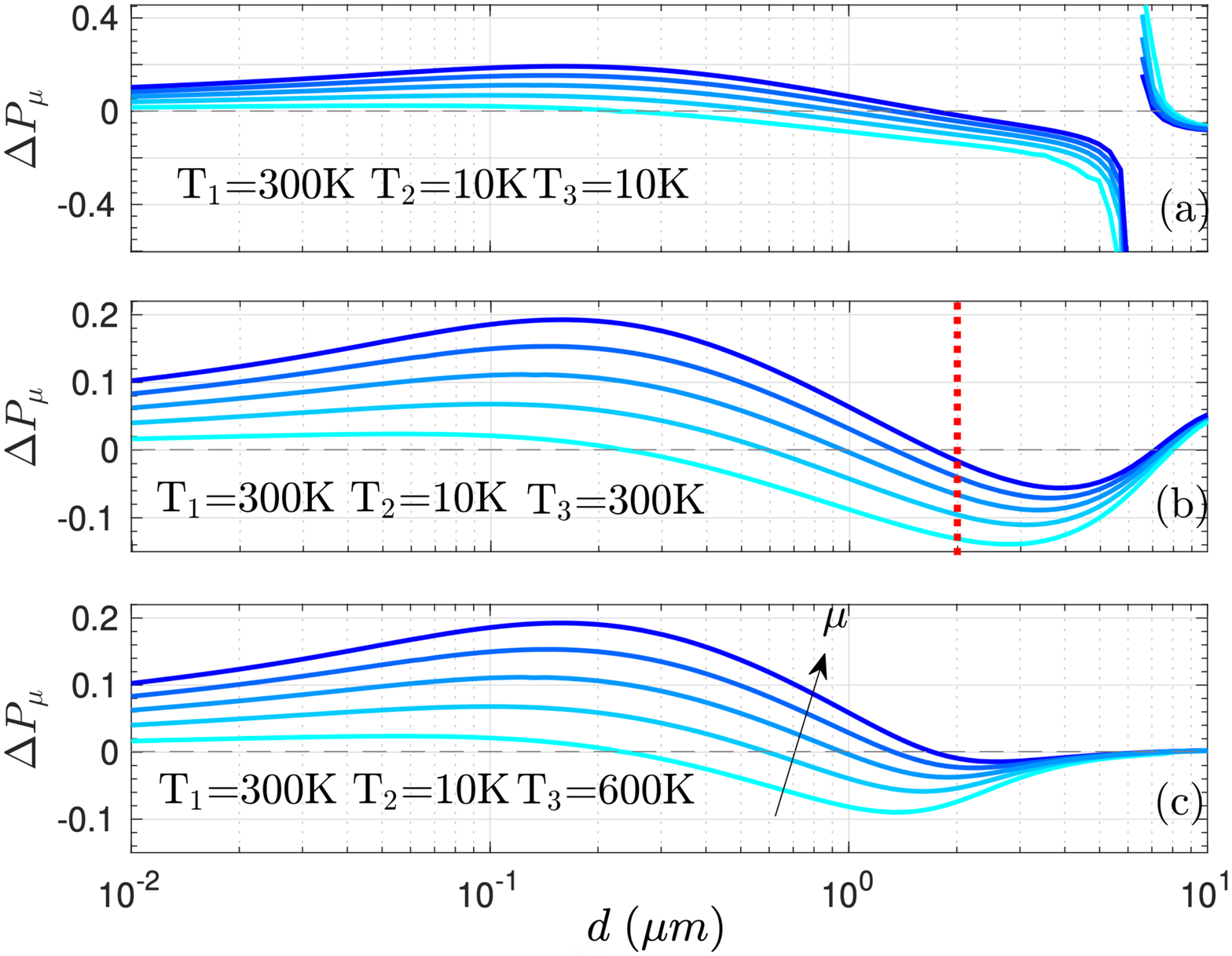}}
\caption{Relative variation $\Delta P_\mu$ of the CLP as a function of the distance between two slab of $SiO_2$ covered with graphene sheets $T_1 \neq T_2$}
\label{Figure15}
\end{figure}

We consider now the case where the two objects are at different temperatures, $T_1 \neq T_2$. Figure \ref{Figure15}(a) displays $\Delta P _\mu$  for different temperatures $T_3$ of the environment. First of all, let us recall that the CLP increases with the chemical potential for suspended graphene sheets if the pressure is attractive. We retrieve this behaviour  in Figs.\ref{Figure15} for distances less than a certain value depending on $\mu$. For example, this distance is about 1.72$\mu$m for $\mu$=1 eV regardless of $T_3$. This behavior is shown in Fig.\ref{Figure16}, where we directly display the CLP for $\mu$ = 0eV and $\mu$=1eV with a zoom around their first intersection. (see sub-panel a).

In the specific case where $T_3$ is fixed at 10 K, we observe that at distances larger than 1.7 $\mu$m, the pressure for $\mu=1$ eV is lower than that for $\mu=0$ eV until a change in sign occurs at around 6 $\mu$m. Subsequently, the pressure for $\mu=1$ eV becomes greater than that for $\mu=0$ eV again, (when $d < 7 \mu$m). Beyond this distance, the pressure for $\mu$ = 1 eV once again becomes lower than that for $\mu$ = 0 eV.

\begin{figure}[ht!]
\centering
{\includegraphics[width=9cm]{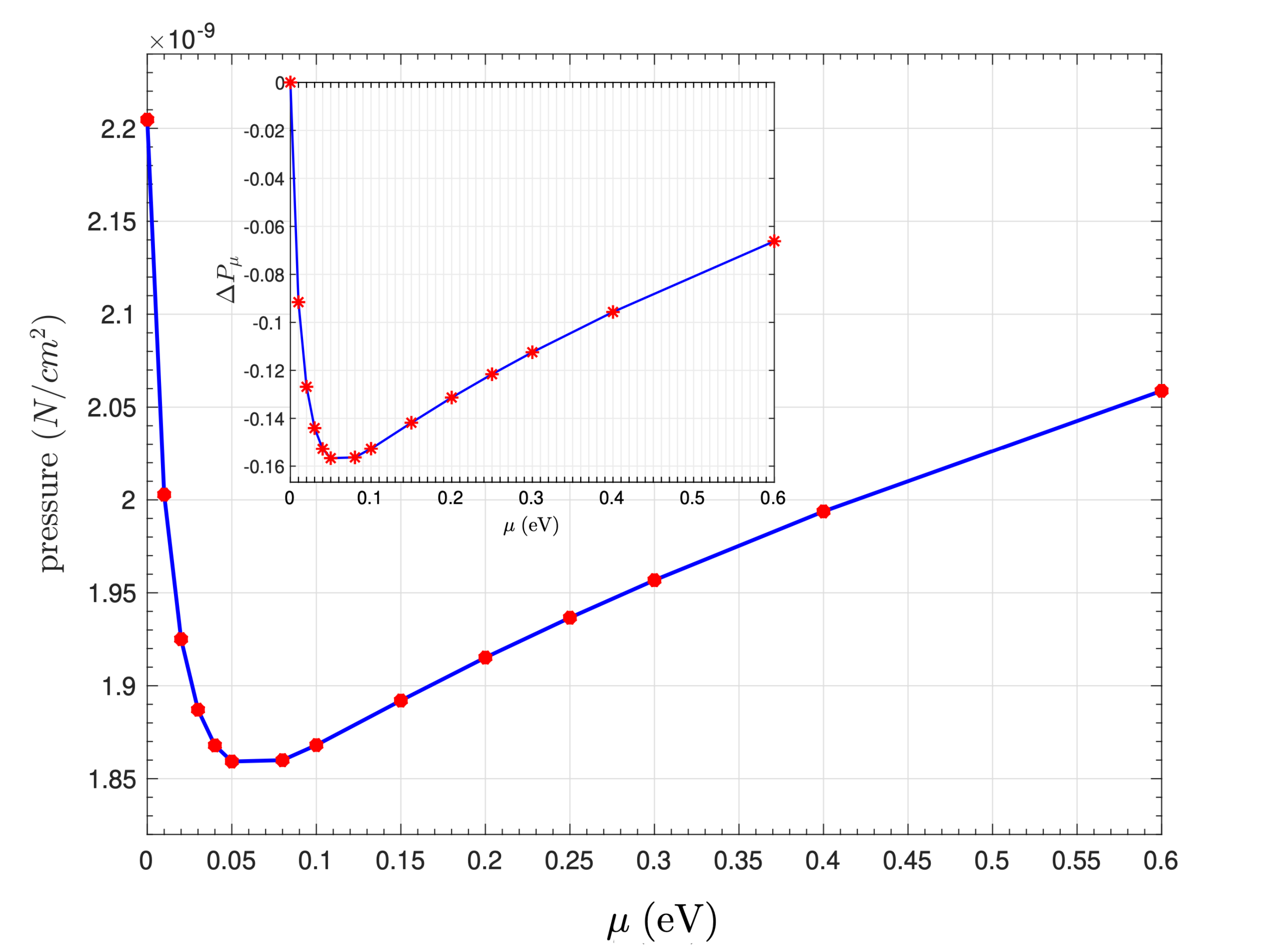}}
\caption{CLP out of thermal equilibrium [Eq. \eqref{Pgeneral}] between two slabs of $SiO_2$ covered with graphene sheets separated by $d$=2$\mu$m and $T_1=300$K, $T_2=10$K and $T_3=300$K }
\label{Figure15_bis}
\end{figure}

\begin{figure}[ht!]
\centering
{\includegraphics[width=8.5cm]{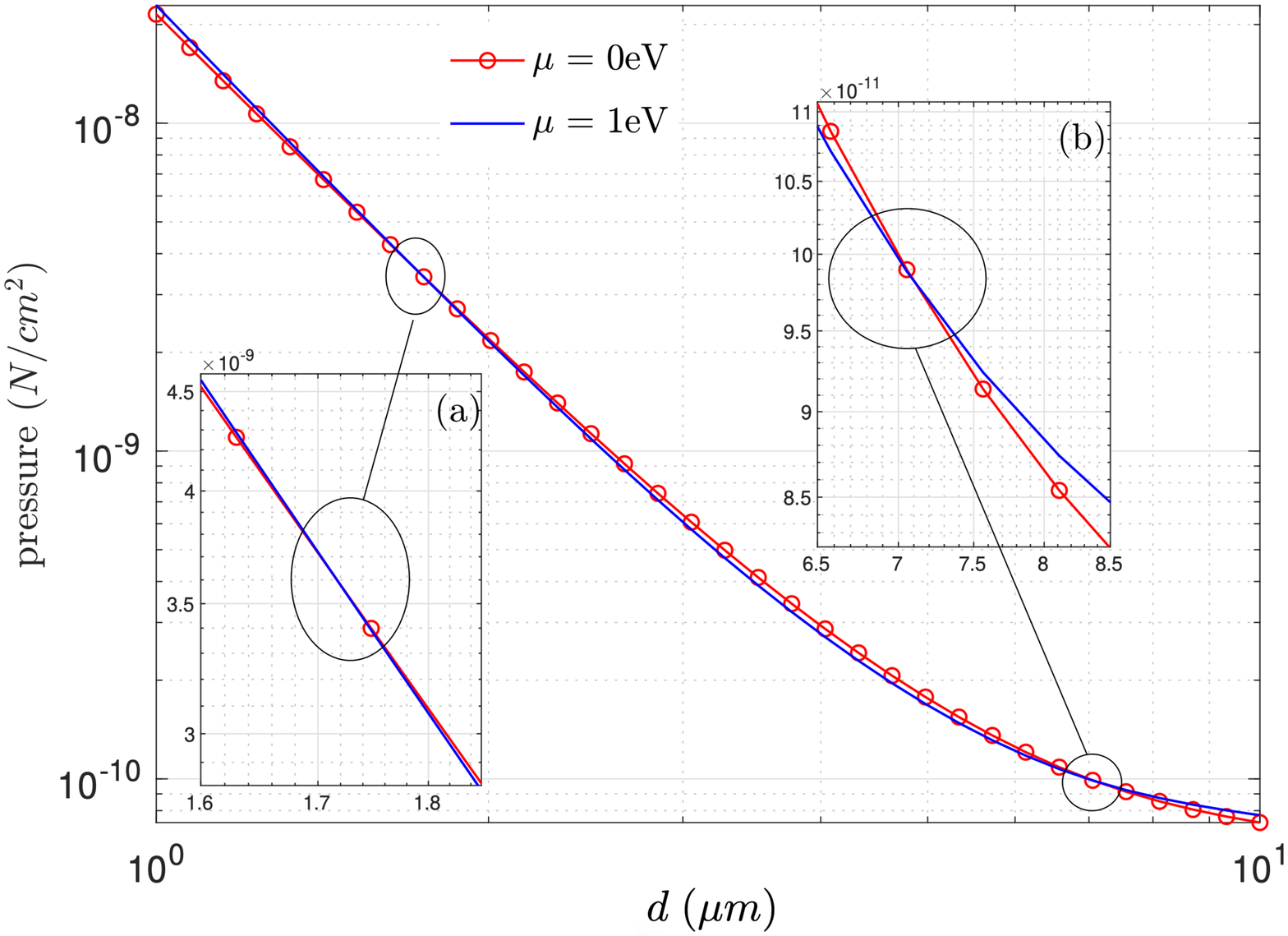}}
\caption{CLP out of thermal equilibrium [Eq. \eqref{Pgeneral}] between two slabs of $SiO_2$ covered with graphene sheets. $T_1=300$K, $T_2=10$K and $T_3=300$K and for $\mu =0$ eV and $\mu =1$ eV.}
\label{Figure16}
\end{figure}

We now shift our attention to the case at $T_3$=300K (see Fig.\ref{Figure15}(b)). It is noteworthy that between 1.7 $\mu$m and 7 $\mu$m, as the chemical potential decreases, the relative variation is expected to approach zero. However, this is not what we observe because the smallest value of $\mu$ we are considering is only 0.2 eV. In order to clarify this point, we focus on a specific distance of 2$\mu$m (red dotted line in Fig.\ref{Figure15}(b)) and gradually decrease the chemical potential towards zero, as shown in Fig.\ref{Figure15_bis} where we display the CLP for different values of $\mu$ and also $\Delta P_\mu$ in the inset. It becomes apparent that this variation is not monotonic. The non-monotonic dependence of the CLP represents a novel feature of non-equilibrium systems that is absent in their equilibrium counterparts. This variation eventually tends towards zero for small values of the chemical potential (below 0.1 eV). After 7 $\mu$m, the opposite occurs, $\Delta P_\mu$ becomes monotonic, and this is clearly shown in sub-panel b of the Fig.\ref{Figure16}.

Finally, for the case of $T_3$=600 K, we observe the same behaviour as in the previous cases, with the pressure for $\mu$=1 eV being lower than that for $\mu$=0 eV after 1.7 $\mu$m. However, as the distance increases, the relative variation tends towards 0, indicating that the difference in CLP between the chemical potentials becomes negligible at larger separations. This behaviour is illustrated in Fig.\ref{Figure15}(c).

When the CLP becomes repulsive, it decreases with chemical potential on the contrary of what happens when it is attractive. This can be directly seen in Fig.\ref{Figure15}(a) for $d>6.2\mu m$.

\section{Conclusions}
We studied the OTE CLP between two graphene layers at two different temperatures in a thermal bath of a third temperature. The graphene layers can be suspended or coated on a fused silica substrate. Unlike the equilibrium case where the pressure is always attractive, the pressure can be tuned to change sign and become repulsive. The repulsive interactions could play a role in alleviating stiction in graphene nanomechanical systems. We also show that the CLP on graphene can be tuned by the chemical potentials. To obtain large changes of the CLP, we find that the temperatures of the graphene sheets need to be kept low in both equilibrium and non-equilibrium situations. While the CLP at equilibrium increases with chemical potential for all distances and temperatures we considered, we find that for OTE CLP the dependence on the chemical potential becomes non-monotonic at certain ranges of distance.  {Finally, one interesting additional degree of liberty that can bring new effect is anisotropy. This is present, e.g., in black phosphorus which has been used to explore the Casimir interaction at thermal equilibrium \cite{Wang:23}}.\\

\section{Acknowledgments}
The work described in this paper was partially supported by a grant "CAT", No. A-HKUST604/20, from the ANR/RGC Joint Research Scheme sponsored by the French National Research Agency (ANR) and the Research
Grants Council (RGC) of the Hong Kong Special Administrative Region.
We acknowledge P. Rodriguez-Lopez for useful discussions.\\

\section*{}
\bibliography{bib_PRB2023}

\end{document}